\documentclass[letterpaper,10pt,conference]{ieeeconf}

\usepackage{graphicx}
\usepackage{amsmath}
\usepackage{amssymb}
\usepackage{dsfont}

\IEEEoverridecommandlockouts
\def\ket#1{|#1\rangle}
\def\bra#1{\langle #1|}

\def\ketbra#1#2{|#1\rangle\langle#2|}

\DeclareMathOperator*{\argmax}{arg\,max}

\title{\LARGE\bf Reinforcement Learning vs.\ Gradient-Based Optimisation for Robust Energy Landscape Control of Spin-1/2 Quantum Networks}

\author{I.\ Khalid \and C.\,A.\ Weidner \and E.\,A.\ Jonckheere \and S.\,G.\ Schirmer \and F.\,C.\ Langbein%
\thanks{I.\ Khalid and F.\,C.\ Langbein are with the School of Computer Science and Informatics, Cardiff University, Wales, UK. {\tt\small KhalidMI@cardiff.ac.uk, frank@langbein.org}.}%
\thanks{C.\,A.\ Weidner is with the Institute for Physics and Astronomy, Aarhus University, Denmark. {\tt\small cweidner@phys.au.dk}.}%
\thanks{E.\,A.\ Jonckheere is with the Department of Electrical Engineering, University of Southern California, Los Angeles, CA 90089. {\tt jonckhee@usc.edu}.}%
\thanks{S.\,G.\ Schirmer is with the College of Science, Swansea University, Swansea, Wales, UK. {\tt s.schirmer@swansea.ac.uk}.}%
}

\begin{document}

\maketitle

\begin{abstract}
We explore the use of policy gradient methods in reinforcement learning for quantum control via energy landscape shaping of XX-Heisenberg spin chains in a model agnostic fashion. Their performance is compared to finding controllers using gradient-based L-BFGS optimisation with restarts, with full access to an analytical model. Hamiltonian noise and coarse-graining of fidelity measurements are considered. Reinforcement learning is able to tackle challenging, noisy quantum control problems where L-BFGS optimization algorithms struggle to perform well. Robustness analysis under different levels of Hamiltonian noise indicates that controllers found by reinforcement learning appear to be less affected by noise than those found with L-BFGS.
\end{abstract}

\section{INTRODUCTION}\label{sec:intro}

Finding robust solutions to Hamiltonian control of quantum devices from superconducting qubits to spintronic circuits to microwave QED to trapped ions~\cite{superconductingqubits, cavityqed, biigspin, trappedions} is crucial to achieve high-fidelity operations in quantum systems that form the building blocks of Noisy Intermediate Scale era Quantum (NISQ) devices~\cite{preskillnisq}. Although early-stage devices are expected to be error-prone and limited in size, they could pave the way to revolutionize computation and simulation at a fundamental level. They have already proven to be effective tools in physically simulating molecular networks~\cite{biigimmanuelblochwork1, biigimmanuelblochwork2, biigimmanuelblochwork3}. Currently, the challenges for NISQ devices are: scalability with system size and robustness to known and unknown uncertainties. For the former, the main problem lies in the exploration of a parameter space growing exponentially in the size of the system, which has been addressed using variational approaches~\cite{variationaleigensolvers, bukov} amongst other work. In this paper we focus on the latter challenge: optimal control with partial observability in the absence of an accurate model, a regime that is particularly challenging for the dominant model-based, open-loop control approaches.

Two frameworks were developed for such problems: dual control theory initiated by Feldbaum in the 1960s~\cite{feldbaum1} and reinforcement learning (RL) for optimal control~\cite{rlforoptimalcontrol}. Both coalesce the control problem to approximate dynamic programming solved using Bellman's principle of optimality~\cite{poo}. Solvers follow the principle of initially exploring and learning the unknown model by probing the system, and, later, exploiting this information for control. Initially the control actions taken by the controlling agent are sub-optimal as it works with a highly uncertain model although they can still be seen as optimal in the sense of solving the Bellman equation step-wise based on the acquired information. Iterated composition of the solutions achieves near optimal solutions, eventually.

Our motivation for RL is to find adaptive model-agnostic ways of performing optimization to solve quantum control problems. These methods, in principle, promise to have less overhead compared with functional variation or Pontyragin-variation-based methods for optimal control which use an analytical model and have been the focus of over half a century of fruitful contribution to quantum control, including algorithms such as GRAPE~\cite{grape} and Krotov~\cite{krotov} that utilise gradient-based optimisation of a model-based target functional. Limited knowledge about the system and control Hamiltonians, and interactions with the environment, however, has a strong effect on the performance of such controls. Sampling over uncertain parameters combined with gradient-based optimisation can find robust controls~\cite{dong1}. Robust, high precision controls have been found by batch optimising neural network target functionals akin to GRAPE~\cite{dong2}.

RL methods are either model-based or model-free, but all methods can in principle be fully model-agnostic. Model-based methods involve creation of a model from scratch, whereas model-free methods skip this step. RL aims to tackle and optimize the trade-off between exploitation and exploration that is the hallmark of dual control. Prior work demonstrated the usefulness of deep RL for quantum optimal control~\cite{dong3} in its application to synthesis of transmon gates~\cite{dalgaard}, coherent transport by adiabatic passage through semi-conductor quantum dots~\cite{coherentcontrol}, and robust two-qubit gmon gate synthesis~\cite{googlerl}.

In this paper we employ RL to find robust quantum controls with a fully model-agnostic approach using single shot measurements, which can be collected experimentally. Instead of passing unitary operators or density matrices to the RL agent, as considered in previous work, we only give it access to experimentally observed data and control parameters~\cite{bukov}. This is in line with real world scenarios where RL may be deployed in an experimental setting with high levels of uncertainty, commonly seen in current setups. We provide a computational resource comparison between policy-gradient-based RL algorithms to motivate our choice of PPO (Proximal Policy Optimisation). We also demonstrate the resilience of RL in finding optimal controllers to measurement and Hamiltonian noise, where analytical methods break down or consume too many resources. Of course analytical model optimization has an advantage over RL when the model describes the physical system well, as no exploration is required. Increasing uncertainties in the model, however, require an RL or exploratory approach. Moreover, although L-BFGS is more likely to find high-fidelity controllers, preliminary robustness analysis for the controllers found by RL suggests that they may be more robust to noise than those found by L-BFGS.

\section{PRELIMINARIES}\label{sec:preliminaries}

\subsection{The Control Problem}

We consider the problem of controlling a spin-1/2 system, described by the XX-Heisenberg spin chain model. Its Hamiltonian is $H_{\text{spin}} = \sum_{m=1,m\neq n}^n J_{mn}\left(X_n X_m + Y_n Y_m\right) + \sum_{n=1}^N{\Delta_n Z_n}$, where $X_n = F_n(\sigma_x) := \mathds{1}^{(1)} \otimes \dots \otimes \mathds{1}^{(n-1)} \otimes \sigma_{x}^{(n)} \otimes \mathds{1}^{(n+1)} \dots \otimes \mathds{1}^{(N)}$ is the expanded Pauli X operator $\sigma_x$ applied on the $n$-th spin in the system; $\Delta=\{\Delta_n\}$ are external control parameters; $J_{mn}$ are the interaction couplings between spin $n$ and $m$. $Y_n$, $Z_n$ are defined equivalently to $X_n$ for the Pauli operators $\sigma_y$ and $\sigma_z$. We only consider a spin chain with uniform nearest-neighbour couplings such that $J_{n,n\pm 1} = 1$ (and all other entries are $0$), which can be thought of as a type of quantum wire. The spin network Hamiltonian $H_{\text{spin}}$ commutes with diagonal operators and therefore the dynamics can be decomposed into excitation subspaces~\cite{TAC2012}. Here, we are only concerned with single excitations, i.e., only one bit of information can propagate through the network at a given time, for simplicity. Therefore we have the single excitation subspace Hamiltonian
\begin{equation}\label{eq:scontrolhamiltonian}
  \hbar^{-1} H_{ss} := \sum_{m\neq n}{J_{mn}\ketbra{m}{n}} + \sum_{n}{\Delta_n \ketbra{n}{n}},
\end{equation}
where $\hbar$ is the reduced Planck constant and $J_{mn}$ and $\Delta_n$ are measured in rad/sec. The unitary time evolution of the single bit propagating through the network is given by the Schr\"odinger Eq., $\frac{d}{dt}\ket{\psi(t)} = -i\hbar^{-1}H_{\text{ss}}(t)\ket{\psi(t)}$, where $\ket{\psi(t)}$ is the $N$ dimensional spin-state vector. This is solved by $\ket{\psi(t)} = \exp(-i\hbar^{-1}H_{\text{ss}} (t-t_0))\ket{\psi(t_0)} = U_{\Delta}(t) \ket{\psi(t_0)}$. The suffix $\Delta$ makes the dependence of the unitary on the control parameters explicit. Consider some target state $\ket{\psi^*}$ and a state propagated by the unitary from an initial state $\ket{\psi(t_0)}$. The state propagation performance is given by the fidelity $\mathcal{F}(\Delta,T) := |\bra{\psi^*}U_{\Delta}(T)\ket{\psi(t_0)}|^2$, measuring how close the propagated and target states are. The resulting optimal control problem is the determination of the control parameters $\Delta, T$ that, e.g., represent the action of applied external magnetic fields, s.t. $\Delta^*, T^* = \argmax_{\Delta,T} \mathcal{F}(\Delta, T)$. We specifically consider transitions between one-hot encoding state vectors (canonical Euclidean basis vectors), consistent with a single bit propagating through the network.

The most common paradigm for quantum control is dynamic~\cite{trainingcat, qcontrolsurvey}, i.e., assuming time-dependent controls, $\Delta_n(t)$, the implementation of which typically requires the ability to rapidly modulate or switch controllers implemented by physical fields (e.g. lasers or magnetic fields). An alternative to dynamic control is time-invariant control, i.e., time-independent control parameters $\Delta_n$~\cite{CDC2015}. This is analogous to shaping the potential landscape to facilitate the flow of information from an initial state to the target state. For example, information encoded in electron or nuclear spins in quantum dots whose potential can be controlled by varying voltages applied to surface control electrodes, creating a potential landscape. The static control problem has fewer parameters, and so is in some sense simpler. Moreover, previous work found evidence concerning good robustness properties of the static controls~\cite{IEEE_TAC_Feedback}. They may also be simpler to implement experimentally as we do not need to modulate control fields, or could be part of a more complex dynamic control scheme. However, the optimisation landscape is challenging~\cite{CDC2015}, and there is no guarantee that the controllers found are robust with respect to uncertainties in the system and interactions with the environment.

\subsection{Reinforcement Learning Control Paradigm}

RL is formulated as a finite Markov decision process (MDP): given an initial state $\mathcal{S}$, a next state $\mathcal{S}'$ can be achieved that carries with it some reward $\mathcal{R}$ by performing some action $\mathcal{A}$. State transitions are assumed to be Markovian and probabilistic and captured by the dynamics model $P(\mathcal{S}', \mathcal{R}| \mathcal{S},\mathcal{A})$, indicating the probability of going from $\mathcal{S}$ to $\mathcal{S'}$ with the action $\mathcal{A}$, gaining $\mathcal{R}$. A trainable policy function
$\pi(\mathcal{A}|\mathcal{S})$ is a non-parametric probability distribution of executing action $\mathcal{A}$ given state $\mathcal{S}$. An RL agent follows $\pi$, interacts with an environment $\mathcal{E}$ and associates a state transition $Y: \mathcal{S} \xrightarrow{\mathcal{A}} \mathcal{S}'$ with a reward function $\mathcal{R}(Y)$. A state-action value function $Q(\mathcal{S}, \mathcal{A})$ or the value function $V(\mathcal{S})=\max_{a}Q(\mathcal{S}, a)$ is learnt via the feedback loop interaction of $\pi$ with $\mathcal{E}$. The environment can be noisy and highly stochastic and yet through the high learning potential of differentiable neural nets as function approximators, a near-optimal $\pi$ or $Q$ can be learnt~\cite{suttonandbarto2018}. Learning $Q$, for example, involves approximately solving the Bellman optimality equation iteratively, as an update rule, at every timestep $k$,
\begin{align}\label{eq:bellmanoptimality}
  Q_k (s, a) &:=
  \mathbb{E}_{\pi}\left[\sum_{k=0}^\infty \gamma^k \mathcal{R}_{\tau+k+1} | S_\tau = s, A_\tau=a \right] && \\ \nonumber
  &\equiv \sum_{s',r}{P\left(s', r| s, a \right)\left[r+\gamma \max_{a'}Q_{k-1} (s', a') \right]}
\end{align}
where $\gamma$ is some future discounting factor and $s',a',r$ are the next state, next action and reward. $Q_k$ is also the expectation over different policy functions $\pi$ of the total discounted rewards obtained from the current timestep onwards.

General theorems for policy and $Q$ (or value) functions guarantee iterated policy improvement~\cite{suttonandbarto2018}. This involves computing a new policy, e.g., $\pi'(s)=\argmax_{a'}Q(s,a')$ for actions $a'$ drawn from some old policy $\pi$. A model is thus not needed for approximately solving the Bellman equation as we can directly optimize over the policy by successively computing better policies (e.g. greedily) to yield an optimal $Q$ function $Q^*(s, a) = \max_{\pi} Q(s,\pi(s))$. For continuous state and action spaces, this approach does not work well. For such high dimensional spaces, we optimize over the policy by making use of the gradient of some expected cumulative performance distribution in terms of the gradient of a differentiable policy $\pi_{\theta}$. Here, $\pi_\theta$ is represented by a linear two-layer neural network with $\theta$ nonparametrically denoting its trainable weights and biases. We assume a similar nonparametric neural network form for $Q$ and/or the value function. Many policy gradient algorithms are based on this idea~\cite{suttonandbarto2018}. Using backpropagation~\cite{backprop} to update $\theta$ in the direction of the policy gradient, we improve the policy $\pi_\theta$. By approximately solving the Bellman Eq.~\eqref{eq:bellmanoptimality} iteratively for finite steps, we evaluate how well it does. Improvement and evaluation are repeated until a convergence criterion is met that we state below.

For our control problem, we define the model agnostic MDP: for learning timestep $\tau$, let $\mathcal{S}_\tau := \{\Delta_{\tau-1}, t_{\tau-1} \} $ and $\mathcal{A}_\tau := \{\delta \Delta_{\tau-1}, \delta t_{\tau-1} \}$ be an action changing $\mathcal{S}_\tau$ by the given values. The reward is $\mathcal{R}_\tau := \mathcal{F}(\ket{\psi_{\tau-1}},\ket{\psi^*})$ where $t_{\tau-1}=T$ is the time for which the Hamiltonian is evolved. The readout time $t_{\tau-1}$ with the $\Delta_{\tau-1}$ are the control parameters for $\pi$ to change such that the reward is improved. Here $t_{\tau-1}$ is the physical readout time, a control parameter, and $\tau$ denotes the algorithmic iteration timestep. Note that this means $\pi$ is a control landscape exploration strategy with the aim to find control parameters that achieve the physical state transition from $\ket{\psi(t_0)}$ to $\ket{\psi^*}$ that maximizes $\mathcal{F}(\Delta, T)$. So the goal, rather than the path to get there, is important, even if of course a shorter path makes finding the goal more efficient. We construct an environment $\mathcal{E}$ that a differentiable policy $\pi_\theta$ can interact with to obtain $(\mathcal{S}_\tau, \mathcal{A}_\tau, \mathcal{R}_\tau)$. The state vector satisfies $\mathcal{S}_\tau = \mathcal{S}_\tau \mod \mathcal{S}_{\text{limit}} $ and we set the the limit $\mathcal{S}_{\text{limit}}$ to be $\pm 10$ for $\Delta_{\tau-1}$ and $30$ for $t_{\tau-1}$ to ensure that the control parameters are physical and realisable in experiments. A reward threshold, e.g. $0.99$, is set as a convergence criterion yielding a single solution vector $\mathcal{S}^*_\tau$ effectively reducing the problem to optimal time-independent Hamiltonian searching. The RL optimization procedure is run for some number of epochs until the reward threshold is achieved. Each epoch consists of a fixed number of timesteps of exploring the landscape from an initial random position. The policy parameters $\theta$ and the $Q$ function are updated via backpropagation every epoch.

The utility of the fact that RL assumes nothing about the analytical form of the model is expected to be useful if the environment $\mathcal{E}$ is stochastic. To test this hypothesis, we consider two noise models: (1) directly augmenting $H_{ss}$ with a structured perturbation $P \sim \mathcal{N}(0,\sigma^2_{\text{noise}})$ where $P$ is a matrix of the same form as $H_{ss}$, i.e. tridiagonal, with normally distributed random values with variance $\sigma^2_{\text{noise}}$ and mean $0$. This simulates noisy or tunably inaccurate physics, e.g. due to leakage of spin couplings. (2) coarse-graining the fidelity $\mathcal{R}_\tau$ to simulate single-shot or inaccurate measurements by replacing it with $\tilde{\mathcal{R}}_\tau \sim \text{Bin}(M, \mathcal{R}_\tau)$, drawn from a binomial distribution where $M$ is the number of measurements made and $\mathcal{R}_\tau$, the true fidelity, is the binomial probability and $\tilde{\mathcal{R}}_\tau$ represents the average single shot measurements to estimate the fidelity probabilities. In this work, the choice of the noise models is motivated purely by generality and simplicity to study control in a learning framework. In the absence of a concrete physical system, we assume all parameters are equally uncertain. For both (1) and (2), correlated noise of a random functional form that actually take into account the physical characteristics of the quantum architecture is also possible and is worth exploring in the context of a particular physical system. Dephasing and decoherence errors that are characteristic of quantum processes are possible to explore under the Sudarshan-Lindbladian evolution of the density matrix~\cite{masterequationlindbladian} and left as future work.

We only consider leakage within the nearest neighbour spins. Another possible source of noise could be leakage to the next nearest neighbours due to cross-couplings between spins in transmon systems or finite laser beam sizes in cold atom or ion systems. For the purposes of this work, however, we neglect next-nearest neighbor coupling as it is negligible or can typically be mitigated in practical systems. We have also made the actions $\mathcal{A}_\tau$ noisy by perturbing the diagonal of $H_{ss}$, but we could have also coarse-grained the actions to account for the finite resolution of the magnetic or laser field that actually implements the controls in a real experiment.

\subsection{Policy Gradient Reinforcement Learning Algorithms}\label{sec:rlalgs}

We try a number of policy gradient RL algorithms to empirically evaluate which one is most suitable for our static control problem: trust region policy optimization (TRPO)~\cite{trpo}, proximal policy optimization (PPO)~\cite{ppo}, deep deterministic policy gradient optimization (DDPG)~\cite{ddpg}, twin policy delayed DDPG (TD3)~\cite{td3} and REINFORCE~\cite{suttonandbarto2018}.

REINFORCE is a pure policy-based algorithm that applies a stochastic gradient ascent update to the policy parameters $\theta \leftarrow \theta + \nabla V_{\pi_\theta}(\mathcal{S}_0)$ for some initial state $\mathcal{S}_0$. The value function gradient is computed using the policy gradient theorem as $\mathbb{E}_\pi \left[\sum_{k=0}^\infty \gamma^k \mathcal{R}_{\tau+k+1} \nabla \pi_\theta / \pi_\theta \right]$ via Monte Carlo sampling over trajectories following $\pi$.

The others are actor-critic algorithms with an acting policy critiqued by $Q_{\pi_\theta}$ or $V_{\pi_\theta}$. The actor-critic methods make use of a replay buffer to store MDP transitions of the form $(\mathcal{S}_\tau, \mathcal{S}_{\tau+1}, \mathcal{A}_\tau, \mathcal{R}_\tau)$ and update $Q_{\pi_\theta}$ or $V_{\pi_\theta}$ following the Bellman update~Eq.~\eqref{eq:bellmanoptimality} by random sampling batches of $\{\mathcal{S}', \mathcal{S}, \mathcal{R}\}$. TD3 and DDPG make use of the deep deterministic policy gradient for $\theta$ updates~\cite{silverdeterministicpgt} and TRPO and PPO use a variant of the natural policy gradient~\cite{kakade2002}. TD3 uses two $Q$ functions and backpropagated updates are in the direction of least change while DDPG employs a vanilla combination of $Q$ and a deterministic policy function jittered with correlated exploration noise. Note that there is no objective constraint on the policy that makes sure it does not vary wildly during parameter updates for different episodes. PPO and TRPO improve upon this by using a KL-divergence constraint between the new and old policy to make sure its variation is constrained during each update. TRPO uses a trust region method~\cite{trustregion} to compute the Hessian of the KL-divergence with a backtracking line search~\cite{linesearch} to update the parameters of the policy. PPO is simpler and uses clipped variation bounds on the KL-divergence that is used directly in the parameter updates of the policy.

\section{RESULTS}\label{sec:results}

\subsection{Cost of Reinforcement Learning Algorithms}\label{ssec:rlcomp}

We first analyse the cost of the policy gradient algorithms from Section~\ref{sec:rlalgs}. The costs are expressed as the number of environment $\mathcal{E}$ calls, corresponding to estimating the fidelity via single-shot measurements, for a run that successfully terminates at a fidelity threshold. This links performance to experimental costs and makes different algorithms comparable without resorting to timing or iteration counts.

\begin{figure}[t]
  \centering
  \includegraphics[width=.85\columnwidth]{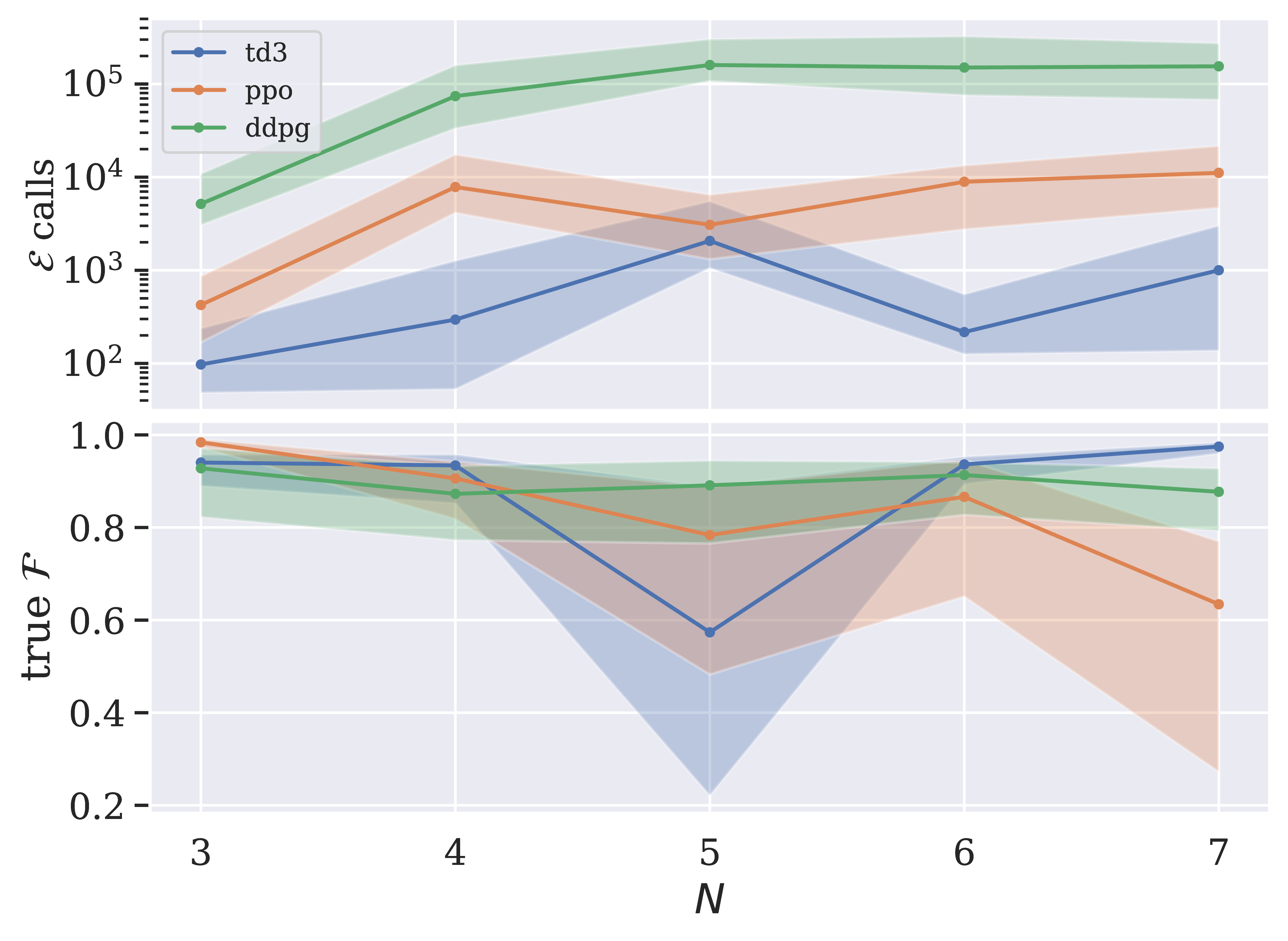}
  \caption{Top: Cost comparison between PPO, TD3 and DDPG for $\ket{0}$ to $\ket{2}$ for chains of length $N=3,\dotsc,7$ with $100$ single shot measurements and $\sigma_{\text{noise}}=0.05$. The algorithms were run $50$ times and the median $\mathcal{E}$ calls are plotted with the interquartile range. A perceived fidelity threshold of $0.99$ was set as the termination criterion. Bottom plot shows true fidelities.}\label{fig:noiseRL}
\end{figure}

We choose to study a noisy transition $\ket{0} \rightarrow \ket{2}$ for chains of length $N=3,\dotsc,7$. We use $100$ single-shot fidelity measurements to estimate the fidelity of a controller and a Hamiltonian perturbation noise of $\sigma_{\text{noise}}=0.05$. The \emph{``perceived'' fidelity} is the stochastic fidelity produced by the noisy environment, as observed from noisy measurements. We compare it to the \emph{``true'' fidelity} of the controller under ideal conditions without noise. A perceived fidelity threshold of $0.99$ is set as termination criterion. Fig.~\ref{fig:noiseRL} shows the median performance of DDPG, PPO and TD3 over $50$ runs. In terms of environment calls, DDPG performs significantly worse compared to PPO and TD3, but it is more difficult to decide between the latter two.

TRPO and REINFORCE were excluded from the study as sufficient statistics could not be obtained. Their behaviour was highly variable and inconsistent due to a lack of successful termination which prevented further analysis. For REINFORCE, we suspect that this was because of the absence of a replay buffer to sample a sufficient variation of transitions and a value/$Q$ function that maps actions to expected rewards to ground policy parameter updates. Similarly, TRPO, while successful in achieving fidelities $> 0.99$ on complicated transitions such as $\ket{0} \rightarrow \ket{3}$ for $N=7$, was algorithmically complex (e.g. the Hessian computation for the KL constraint) and took much longer than the rest.

\subsection{Robustness of Reinforcement Learning Controllers}\label{ssec:mcrarl}

The robustness of the controllers found by RL in Section~\ref{ssec:rlcomp} remains unclear and serves as a further criterion to choose a suitable RL algorithm. We conduct a Monte Carlo robustness analysis (MCRA) using variable Hamiltonian perturbation noise $\sigma_{\text{noise}}$ of the $50$ controllers computed for each chain length for all three algorithms. For each controller $\mathcal{S}_\tau$ found, we perturb the Hamiltonian $\mathcal{H}_{ss}$ using noise of the same triagonal form with mean $0$ and the variance $\sigma^{(i)}_{\text{noise}} = 0.1 k/9$, $k=0,\dotsc,9$. We then evaluate the true fidelities $\mathcal{F}$ of the controller $\mathcal{S}_\tau$ for each level of perturbation without any additional noise. We repeat this ten times for all $50$ controllers and combine the results into a single fidelity distribution. This allows us to judge the expected fidelity of the controllers found by the algorithm.

\begin{figure}[t]
  \centering
  \includegraphics[width=.9\columnwidth]{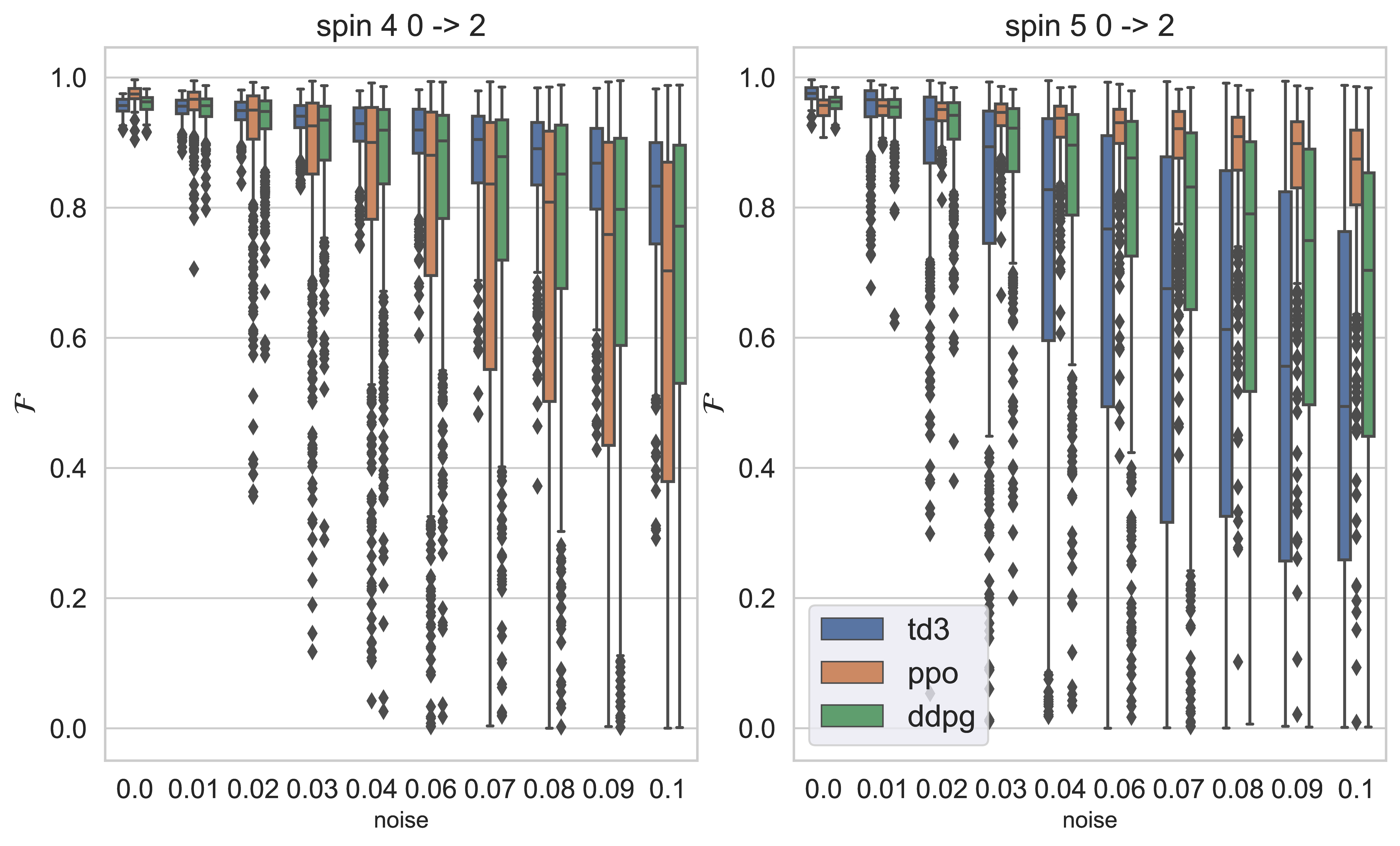}
  \caption{Robustness analysis for PPO, TD3 and DDPG for $\ket{0}$ to $\ket{2}$ for the $50$ controllers found in Section~\ref{ssec:rlcomp} for chains of length $N=4$ (left) and $5$ (right). Ten levels of perturbation noise $\sigma_{\text{noise}}=0,\dotsc,0.1$ are considered for each controller which is evaluated ten times to yield $500$ points per box-plotted fidelity distribution.}\label{fig:noiseRLcontroller}
\end{figure}

The distributions are represented non-parametrically as 1D box-plots as shown in Fig.~\ref{fig:noiseRLcontroller} (the other cases are similar, but are omitted due to space limitations). This figure highlights that some fidelity distributions are heavy tailed with many outliers, meaning there is significant variation of fidelity between some controllers under perturbation. DDPG controllers, despite making more function calls, were the least robust when it came to preserving the interquartile width of the performance distribution. For PPO vs.\ TD3, there are cases where TD3 is better than PPO's and vice versa. However, PPO's performance was more consistent compared with TD3's. TD3, similar to REINFORCE and TRPO, showed a high variation in successful termination, getting stuck indefinitely at local minima for some problems, and there were gaps in the collected statistics due to timeouts. So we were only able to collect statistics for some $N$ for some of the cases in Section~\ref{ssec:rlcomp} without rerunning multiple times. On balance, we find that PPO performs most consistently compared to the other RL algorithms for multiple repetitions for different spin transitions. Therefore, we decided to focus on PPO for the comparison with gradient-based optimisation.

\subsection{Cost of PPO vs.\ L-BFGS}\label{ssec:qnewton}

Even though PPO is not conclusively better from these results, we chose PPO as the single algorithm for comparison with gradient-based optimization methods as we found it (1) faster for data collection to get enough statistics from multiple training runs and (2) sufficient to empirically represent the class of policy gradient algorithms for our problem. TD3 and DDPG might also be suitable for the study but were not pursued chiefly due to time constraints.

A first step to compare PPO with gradient-based optimisation is to analyse the costs in terms of number of $\mathcal{E}$ calls (see Section~\ref{ssec:rlcomp}) under the noiseless dynamics of the ideal model. For gradient-based optimisation, we use L-BFGS with restarts, which performed well on the studied control problem in earlier work~\cite{CDC2015}.

\begin{figure}[t]
  \centering
  \includegraphics[width=.95\columnwidth]{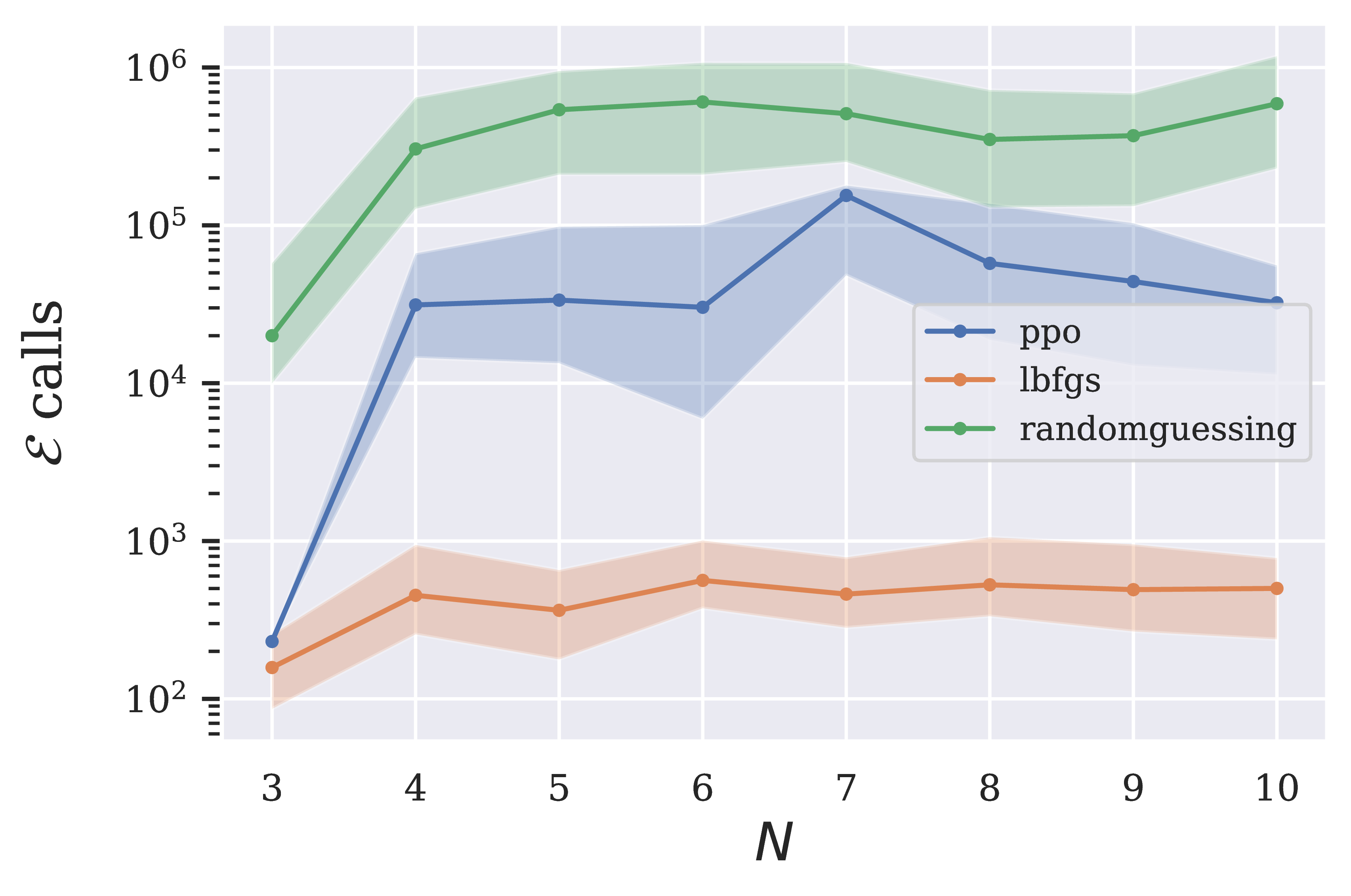}
  \caption{Comparision between L-BFGS, PPO and randomly guessing controllers for $\ket{0}$ to $\ket{2}$ for chains of length $N=3$ to $N=10$ without noise. The algorithms were run $50$ times and the median $\mathcal{E}$ calls are plotted with the interquartile range. A threshold of $\mathcal{F}=0.99$ is set for termination.}\label{fig:nonoise}
\end{figure}

Fig.~\ref{fig:nonoise} shows how function calls scale with the length of the spin chain, $N=3,\dotsc, 10$, for a transition $\ket{0}$ to $\ket{2}$ for PPO, L-BFGS and randomly guessing controllers. The randomly guessed controllers are used to benchmark potential deviations in the computational difficulty of the problem. We stop once a fidelity threshold of $0.99$ is crossed. The spin chain transition is computationally similar for all $N$ as it depends largely on the relative distance between the spins, the control and time constraints, which are kept constant for all the problems we study. There is an initial jump from $N=3$ after which all algorithms manifest a quite flat increase in the number of function calls as the length of the chain increases. This is likely because transitions in the $3$-chain are easier to achieve as simple Rabi oscillations which are generally trap free, and due to the existence of analytical solutions for this case which are absent for longer chains.

\begin{figure}[t]
  \centering
  \includegraphics[width=.9\columnwidth]{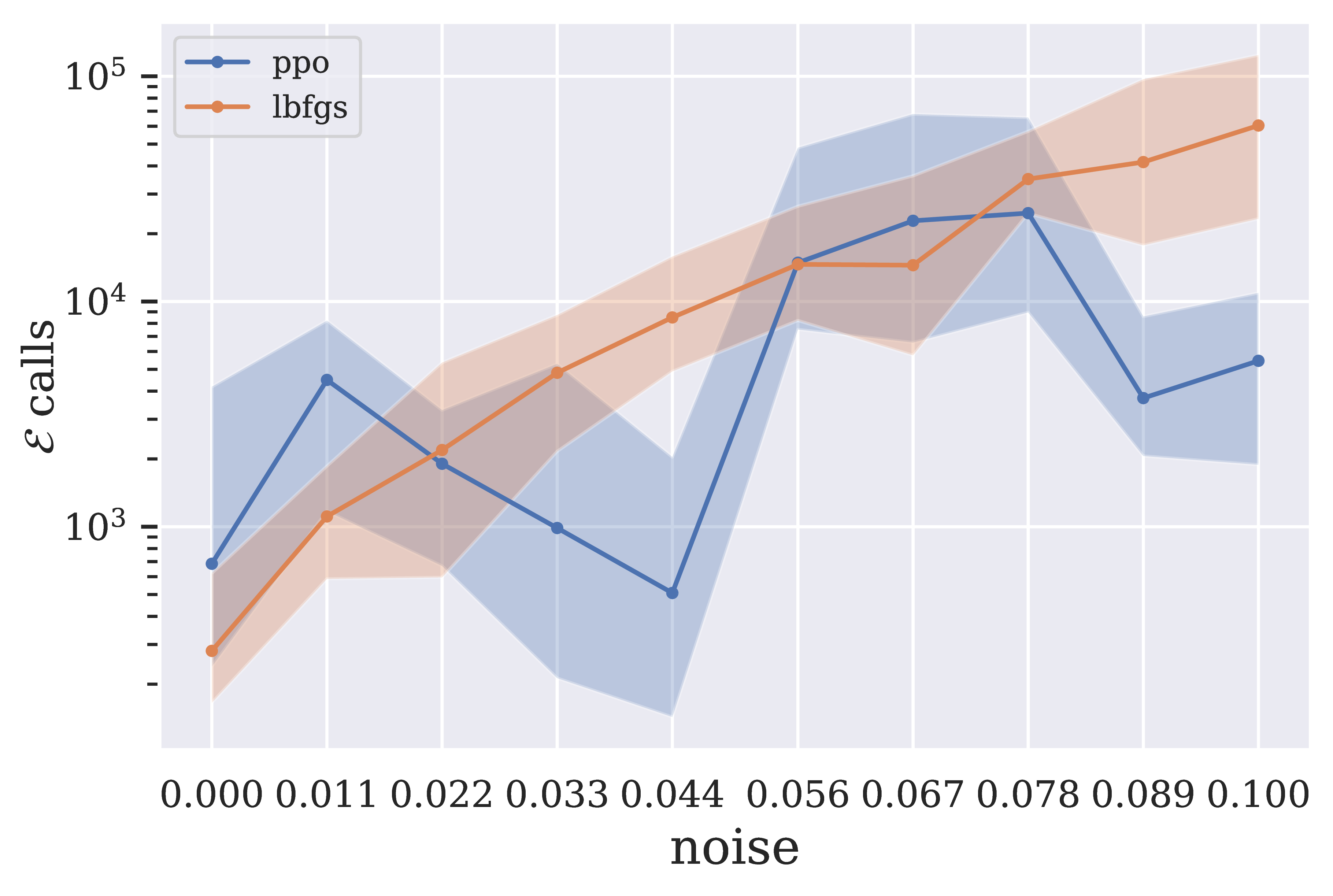}
  \caption{Number of $\mathcal{E}$ calls comparison between L-BFGS and PPO for $\ket{0}$ to $\ket{2}$ for a chain of length $N=4$ as a function of Hamiltonian perturbation noise $\sigma_{\text{noise}}$ with a termination fidelity threshold of $0.98$. The algorithms were run $50$ times and median $\mathcal{E}$ calls are plotted with interquartile range.}\label{fig:noisylbfgs}
\end{figure}

It is not surprising to observe that for an accurate model L-BFGS is mostly two orders of magnitude better than PPO. PPO has to consume most of the calls to build up an internal representation of the model before it can start optimizing. Adding small stochastic noise to the Hamiltonian should degrade the performance of L-BFGS considerably in terms of the number of function calls. To analyze this, we relax the termination constraint on fidelity to $0.98$ and consider only perturbations to $H_{\text{ss}}$ without single shot measurement noise. Single-shot measurement or perturbation noise renders L-BFGS incapable of estimating fidelities over $0.99$ without making many millions of function calls (hence the reduction to $0.98$). Fig.~\ref{fig:noisylbfgs} demonstrates an approximately exponential rise in $\mathcal{E}$ calls for L-BFGS as the strength of the perturbation $\sigma_{\text{noise}}$ is increased from $0$ to $0.1$. Clearly Hamiltonian perturbations deteriorate the performance of L-BFGS, while PPO keeps performing on a similar level than without noise.

Large fluctuations for PPO at certain noise levels likely imply that it is unable to find robust solutions there. The fluctuations may be linked to the noise level and the existence of an optimal noise level at which highly robust solutions can be found. More work, however, is needed to test this idea.

Single shot measurement noise considered in Section~\ref{ssec:rlcomp} has not been employed here, as this would have made the problem even harder for L-BFGS as it has not been designed for noisy optimisation. Overall these results are likely due to high sensitivity of the optimization descent step of L-BFGS to small perturbations in the low rank Hessian components. This causes the number of iterations to steeply increase. Note that $\mathcal{E}$ calls go down for PPO from around $10^5$ to around $10^4$ in Fig.~\ref{fig:nonoise}, and we observe a similar effect in Fig.~\ref{fig:noiseRL}.

\subsection{Robustness of PPO and L-BFGS Controllers}\label{ssec:mcrarlqnewton}

\begin{figure}[t]
  \centering
  \includegraphics[width=.95\columnwidth]{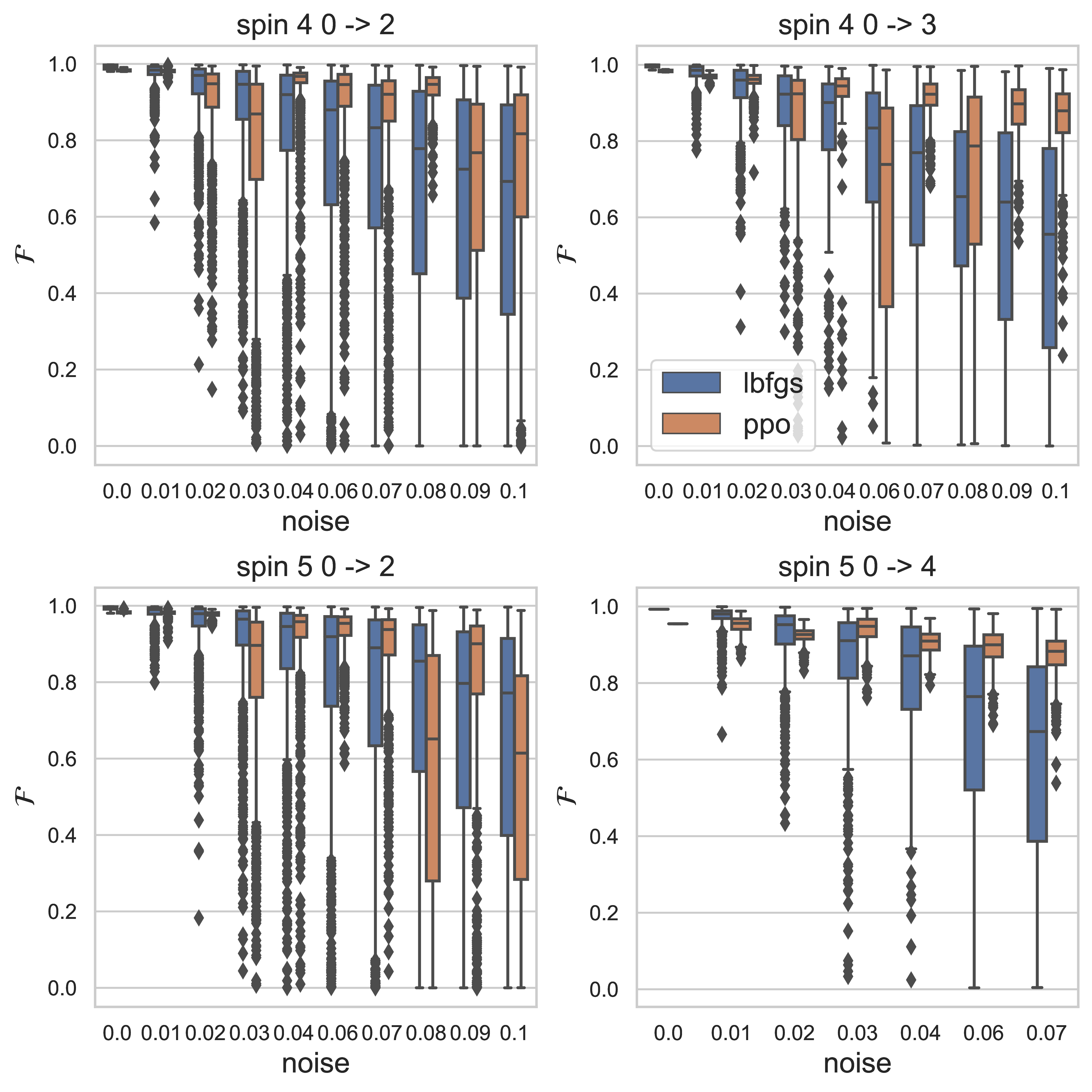}
  \caption{Comparison of $100$ L-BFGS controllers computed without noise and $100$ PPO controllers trained under Hamiltonian perturbation noise $\sigma_{\text{noise}}$ for transitions to the middle and end of chains of length $N=4,5$.}\label{fig:bestpponotfixed}
\end{figure}

We conduct an MCRA (see Section~\ref{ssec:mcrarl}) to compare robustness of $100$ controllers found by L-BFGS under ideal conditions and model-free PPO under low Hamiltonian perturbation noise. There are two cases worth considering: (1) Robustness of PPO controllers found at different levels of Hamiltonian perturbation; (2) The robustness of PPO controllers w.r.t.\ Hamiltonian perturbation found at a particular noise level. Both cases are compared to $100$ L-BFGS controllers for each transition using the ideal model without noise. The termination condition, in all cases, is $\mathcal{F}\geq0.99$.

For (1), we consider transitions to the middle and end for $N=4,5$, as shown in Fig.~\ref{fig:bestpponotfixed}. We use PPO controllers trained with Hamiltonian perturbation noise $\sigma_{\text{noise}}$ that corresponds to the noise level on the $x$ axis from $0.01$ to $0.1$. We find, as expected, that the width of the fidelity distribution for L-BFGS controllers slowly increases as $\sigma_{\text{noise}}$ is increased from $0$ to $0.1$. The expected fidelity is further dropping from being concentrated around $\mathcal{F}=0.99$ to a very flat width and increasingly heavier tail, down to $\mathcal{F}=0$. For PPO controllers, however, we observe that at certain noise levels, e.g., $\sigma_{\text{noise}}=0.01,0.04,0.07$, the controllers found for all problems have narrow distributions compared with L-BFGS. At other noise levels, e.g., $\sigma_{\text{noise}}=0.08,0.1$ for $N=5, \ket{0}$ to $\ket{2}$, they have wider distributions for some problem, but also narrow distributions for others, e.g., $\sigma_{\text{noise}}=0.08,0.1$ for $N=4, \ket{0}$ to $\ket{2}$. We conjecture that added structured perturbations may have a smoothing effect on the optimization landscape which would result in either filtration or creation of ``barriers'' near optima in some cases.

For (2), we consider in addition to the cases of (1), also transitions to the middle for $N=6,7$. Here the PPO controllers have been computed for low Hamiltonian perturbation noise $\sigma_{\text{noise}}=0.01$. Both the L-BFGS controllers and the PPO controllers become worse with increasing noise levels. However, the PPO controllers drop off slower, except in the case of $N=6$, $\ket{0}$ to $\ket{3}$. This suggests that overall PPO is more likely to find robust controllers.

\begin{figure}[t]
 \centering
 \includegraphics[width=.95\columnwidth]{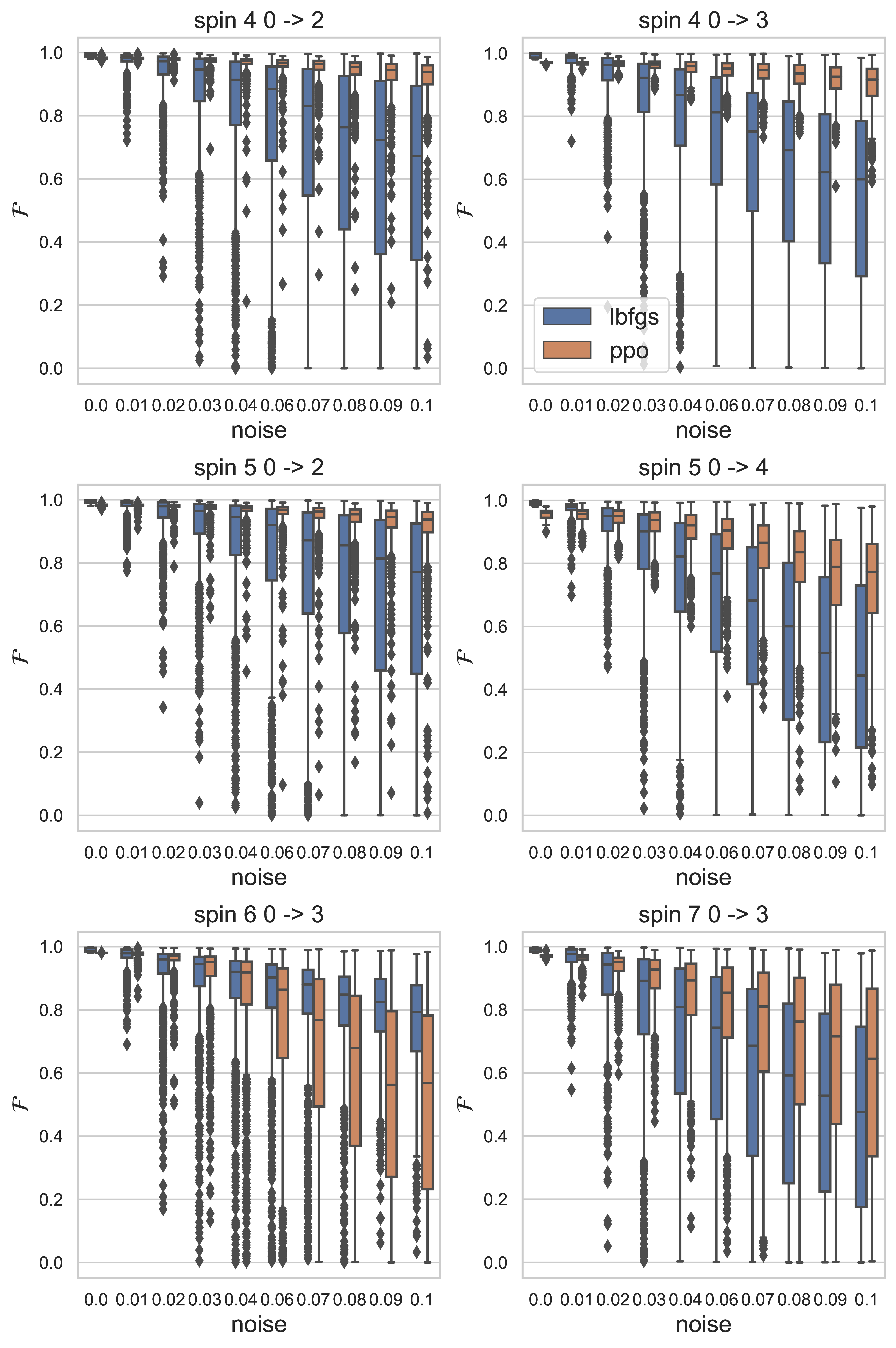}
 \caption{Comparison of controllers found by L-BFGS without noise and PPO trained under low Hamiltonian perturbation noise $\sigma_{\text{noise}}=0.01$ and perfect measurements. We consider transitions to the middle and end of chains of length $N=4,5$ and to the middle for $N=6,7$.}\label{fig:bestppofixed}
\end{figure}

\begin{figure}[t]
  \centering
  \includegraphics[width=.93\columnwidth]{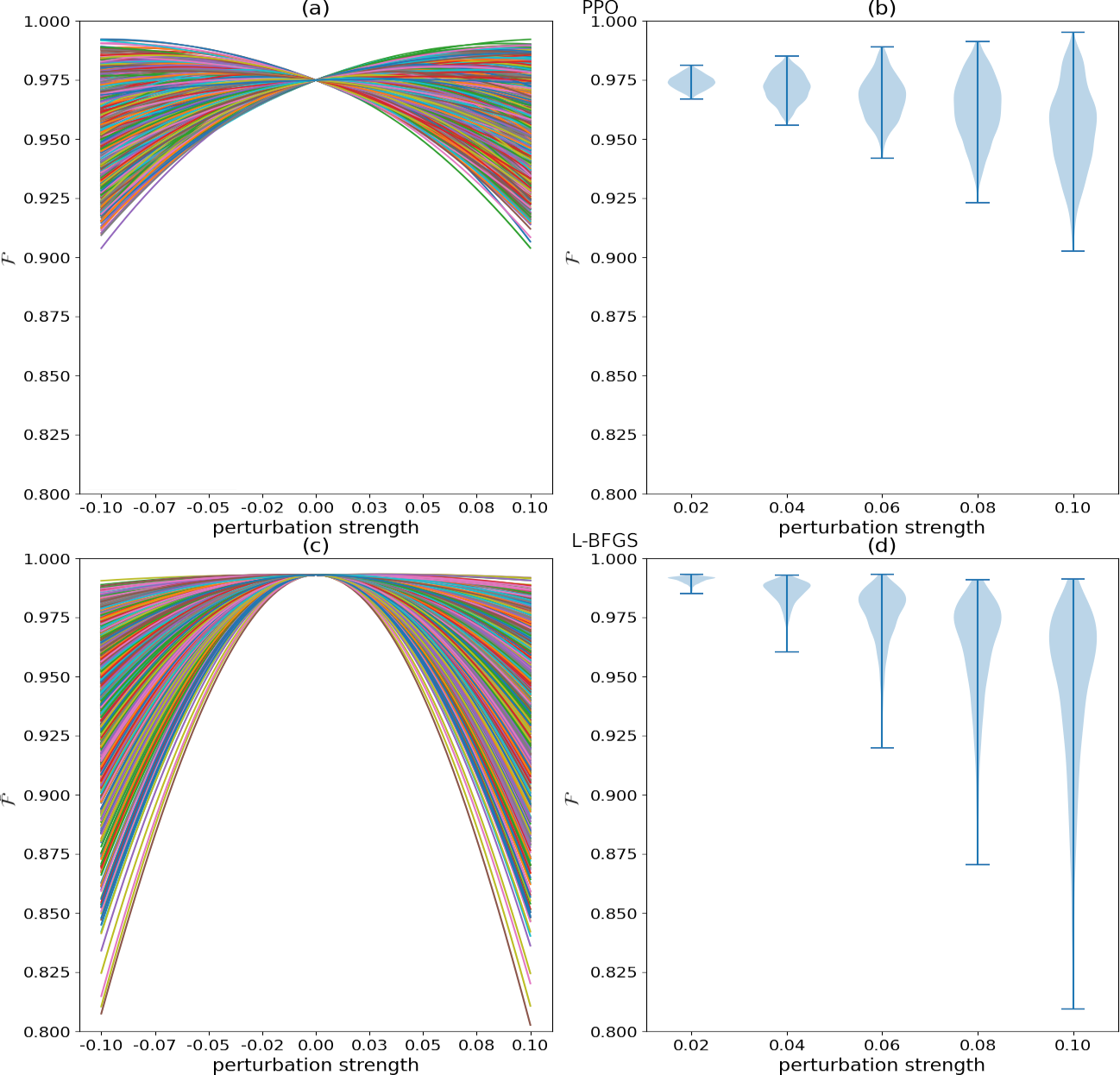}
  \caption{Robustness comparison of a well performing PPO (top) and L-BFGS (bottom) controller for $N=5$, $\ket{0}$ to $\ket{4}$. (a) and (c) show $1,000$ fidelity curves, sampled along different Hamiltonian perturbation directions; (b) and (d) show density distributions of these curves at the\,perturbation strengths.}\label{fig:ring5}
\end{figure}

To investigate this further, the performance of a well-performing PPO and L-BFGS controller for the $N=5$, $\ket{0}$ to $\ket{4}$ transition is compared. For each algorithm, we select the controller with the the highest median fidelity across the ten noise levels to account for the heavy tail nature of the performance distribution. The Hamiltonian is perturbed as $H_{ss} + \delta P$ where $P$ is the perturbation direction and $\delta$ its strength. $P$ is sampled uniformly on a $9D$ Euclidean sphere, created by the five perturbation for $\Delta_n$ and a further four for the coupling strengths. The fidelity was computed along these directions for $\delta$ from $-0.1$ to $0.1$. The density of the curves is estimated at specific perturbation strengths and plotted (see Fig.~\ref{fig:ring5}). The PPO controller is clearly not at a fidelity maximum, so some perturbations have a chance to improve the fidelity. The L-BFGS controller is at a fidelity maximum, which means that most perturbation directions, including those on the couplings which are not control parameters, reduce the fidelity. Similar behaviour has been observed for other controllers.

\section{DISCUSSION AND CONCLUSION}\label{sec:discussion}

Our main finding is that policy gradient RL methods allow nonparametric constructions of optimization models even under highly noisy conditions as seen in Section~\ref{ssec:rlcomp} where pure model-based methods perform poorly as seen in Section~\ref{ssec:qnewton}. We have quantified costs in terms of the number of function or environment calls. In the absence of noise, RL performance is lower bounded by model-based optimisation and upper bounded by pure random guessing. This implies that a nonparametric model is being constructed. The cost of model construction is relatively bounded by random guessing for RL under noisy conditions. However, the number of calls is still high. Model-based RL or Bayesian methods could be explored to reduce the reliance on information acquisition.

In Section~\ref{ssec:mcrarl}, a Monte Carlo robustness analysis and consistency of PPO for variations of the energy landscape control problem is used to motivate our choice of PPO for comparison with L-BFGS with restarts to understand robustness of controllers found by RL. We demonstrate that RL controllers found under low Hamiltonian perturbation noise levels are typically more robust compared with those found by L-BFGS but there is variation within the quality of their robustness that needs to be explored more as a function of their clustering and correlation of locations in the optimization landscape. It appears that in some cases RL finds controllers that may not be optimal for the ideal model, but perform robustly at high fidelity under noisy conditions. This suggests that Hamiltonian noise in particular can improve robustness of some controllers. RL is a promising avenue for feedback adaptive control with less overhead compared with variational methods and is arguably comparatively better with uncertainties. However, a careful construction of the control problem in an RL paradigm is needed before its application.

\section{ACKNOWLEDGMENTS}

I.\,K.\ is supported by a Cardiff University PhD scholarship. E.\,A.\,J.\ is supported by NSF IRES 1829078 grant.

\end{document}